\documentclass[]{spie}  

\usepackage{amsmath,amsfonts,amssymb}
\usepackage{graphicx}
\usepackage[colorlinks=true, allcolors=blue]{hyperref}
\usepackage{siunitx}
\usepackage{placeins}
\usepackage{multicol}
\usepackage{multirow}
\usepackage{array}
\usepackage{booktabs}
\usepackage{graphicx}
\usepackage{setspace}
\usepackage{tocloft}
\usepackage{subcaption}
\usepackage{wrapfig}

\title{X-ray speed reading with the MCRC: prototype success and next generation upgrades}

\author[a]{Peter Orel}
\author[a,b]{Abigail Y. Pan}
\author[a]{Sven Herrmann}
\author[a]{Tanmoy Chattopadhyay}
\author[a]{Glenn Morris}
\author[a,b]{Haley Stueber}
\author[a,b,c]{Steven W. Allen}
\author[a]{Daniel Wilkins}
\author[d]{Gregory Prigozhin}
\author[d]{Beverly LaMarr}
\author[d]{Richard Foster}
\author[d]{Andrew Malonis}
\author[d]{Marshall W. Bautz}
\author[e]{Michael J. Cooper}
\author[e]{Kevan Donlon}

\affil[a]{Kavli Institute for Particle Astrophysics and Cosmology, Stanford University, 452 Lomita Mall, Stanford, CA 94305, USA}
\affil[b]{Department of Physics, Stanford University, 382 Via Pueblo Mall, Stanford, CA 94305, USA}
\affil[c]{SLAC National Accelerator Laboratory, 2575 Sand Hill Road, Menlo Park, CA 94025, USA}
\affil[d]{MIT Kavli Institute for Astrophysics and Space Research}
\affil[e]{MIT Lincoln Laboratory}

\authorinfo{Further author information: (Peter Orel)\\P.O.: E-mail: porel@stanford.edu, Telephone: +1 737 230 2800}

\begin{document} 
\maketitle

\begin{abstract}
The Advanced X-ray Imaging Satellite (AXIS) is a NASA probe class mission concept designed to deliver arcsecond resolution with an effective area ten times that of Chandra (at launch). The AXIS focal plane features an MIT Lincoln Laboratory (MIT-LL) X-ray charge-coupled device (CCD) detector working in conjunction with an application specific integrated circuit (ASIC), denoted the Multi-Channel Readout Chip (MCRC). While this readout ASIC targets the AXIS mission, it is applicable to a range of potential X-ray missions with comparable readout requirements. Designed by the X-ray astronomy and Observational Cosmology (XOC) group at Stanford University, the MCRC ASIC prototype (MCRC-V1.0) uses a 350 nm technology node and provides 8 channels of high speed, low noise, low power consumption readout electronics. Each channel implements a current source to bias the detector output driver, a preamplifier to provide gain, and an output buffer to interface directly to an analog-to-digital (ADC) converter. The MCRC-V1 ASIC exhibits comparable performance to our best discrete electronics implementations, but with ten times less power consumption and a fraction of the footprint area. In a total ionizing dose (TID) test, the chip demonstrated a radiation hardness equal or greater to 25 krad, confirming the suitability of the process technology and layout techniques used in its design. The next iteration of the ASIC (MCRC-V2) will expand the channel count and extend the interfaces to external circuits, advancing its readiness as a readout-on-a-chip solution for next generation X-ray CCD-like detectors. This paper summarizes our most recent characterization efforts, including the TID radiation campaign and results from the first operation of the MCRC ASIC in combination with a representative MIT-LL CCD.  
\end{abstract}


\section{Introduction}
\label{sec:intro}  

Charged-coupled device (CCD) X-ray imaging detectors transformed the field of X-ray astronomy, offering simultaneous imaging and moderate spectroscopic capabilities over the $\sim$ 0.3 - 10 keV energy range. These detectors served to unlock a vast array of scientific discoveries, from precise measurements of the thermodynamic properties of the hot, X-ray emitting plasma in groups and clusters of galaxies to detailed measurements of matter in its final moments before plunging into supermassive black holes. To enable future, strategic X-ray missions, the 2020 Physics of the Cosmos Program Annual Technology Report (PATR) identified ``fast, low-noise, megapixel X-ray imaging arrays" as a top-priority technology development need for future strategic astrophysics missions. One such X-ray mission is the proposed Advanced X-ray Imaging Satellite (AXIS). AXIS, a response to NASA’s Astrophysics Probe Explorer (APEX) Announcement of Opportunity, will be a critical part of the 2030s astrophysics landscape. With 1-2 arcsecond quality imaging across a 24 arcminute diameter field-of-view and a sensitivity greater then 10$\times$ that of the Chandra X-ray observatory at launch (Chandra's soft X-ray sensitivity has decreased notably over time), AXIS offers X-ray capabilities that are well-matched in depth to the James Web space telescope and other near-term and future ground- and space-based facilities that will dominate the next 15 years of astronomical research \cite{AXISpaper1}. 

\begin{figure}
\begin{center}
\begin{tabular}{c} 
\includegraphics[width=6.5in]{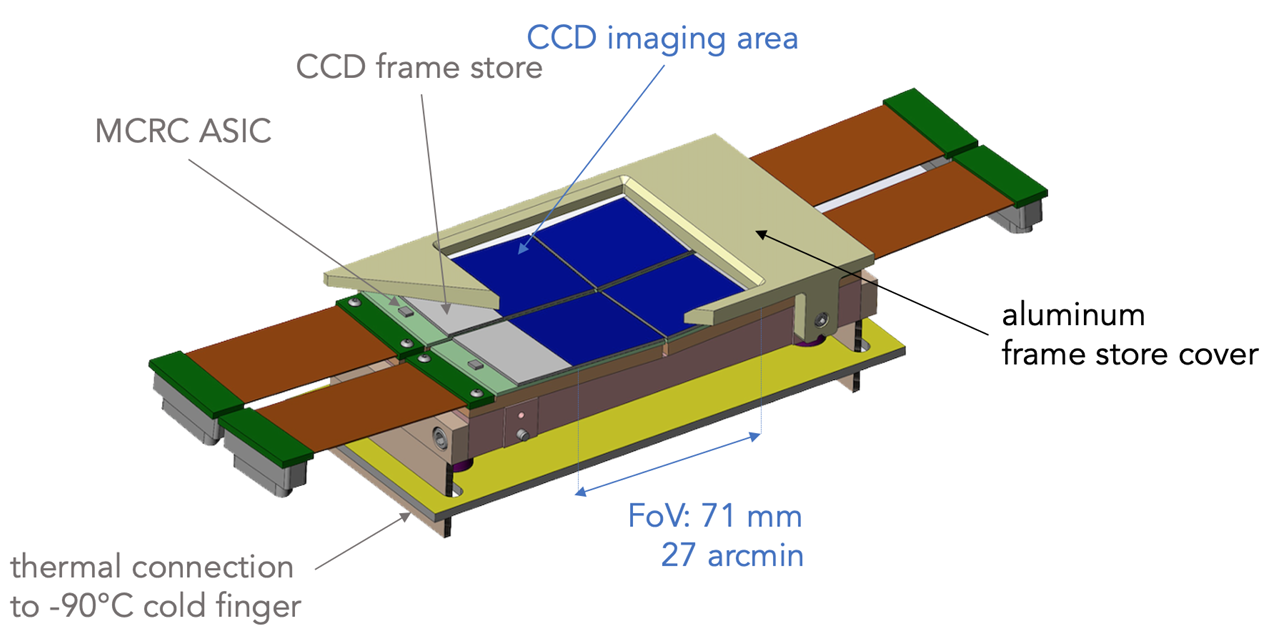}
\end{tabular}
\end{center}
\caption{Concept of the AXIS X-ray imager focal plane array. The focal plane is composed of four sub-modules, each consisting of a frame-transfer CCD with 1440 x 1440 pixels of 24 \unit{\micro m} size, resulting in a X-ray imager with a total of 8 Mega-pixels. The field of view is $27\times27$ arcmin$^2$, with each pixel corresponds spanning $0.5\times0.5$ arcsec$^2$ on the sky. The nominal operation temperature is 183 K (-90 \unit{\degree C}).}
\label{fig:axis_fpa}
\end{figure}

The high-speed AXIS X-ray camera incorporates a focal plane array with CCDs designed and fabricated by MIT Lincoln Laboratory (MIT/LL). The CCDs have multiple outputs to increase the data rate from the 1440 $\times$ 1440 pixel imaging area, and are coupled with an application-specific integrated circuit (ASIC) specifically designed for compact, low-power consumption, high-speed, and low-noise readout. Conceived by the XOC group at Stanford University, the ASIC is called the Multi-Channel Readout Chip (MCRC). Figure \ref{fig:axis_fpa} shows the AXIS focal plane array, which consists of four fast-readout frame-store CCDs arranged in a 2$\times$2 pattern, each with a dedicated MCRC readout chip co-mounted on the interposer. In addition, an aluminum cover shields the frame store regions from the focused celestial X-rays during readout, and provides the ASICs with additional particle damage protection. The entire array is maintained at 183 K using a cold finger connection and active trim heaters. AXIS will be placed in a low-inclination ($<$ 8 \unit{\degree}), low-Earth orbit (LEO). The primary drivers for this choice are the need to maintain rapid, on-demand (low data-rate) communications for time-domain science, and to obtain a low radiation environment, providing a low detector background and slowing detector degradation \cite{AXIShihspeedcamera}. 

Notwithstanding shielding and LEO orbit placement, the readout electronics still needs to be radiation tolerant. As such, the MCRC prototype (MCRC-V1) has been exposed to gamma radiation and its total ionizing dose (TID) response has been measured and is reported in section \ref{sec:mcrc_and_rad00}. Further test results for the MCRC-V1 prototype in both a stand alone configuration and paired with prototype AXIS CCD detectors are presented in section \ref{sec:workingwithmcrc}, demonstrating the technology readiness. Future improvements and updates that will be incorporated in the second generation of the MCRC chip are discussed in section \ref{section:conclusions}.     

\section{Architecture and design}
\label{sec:mcrcarch}

\begin{figure}
    \centering
    \begin{subfigure}{.64\textwidth}
    \includegraphics[width=\linewidth]{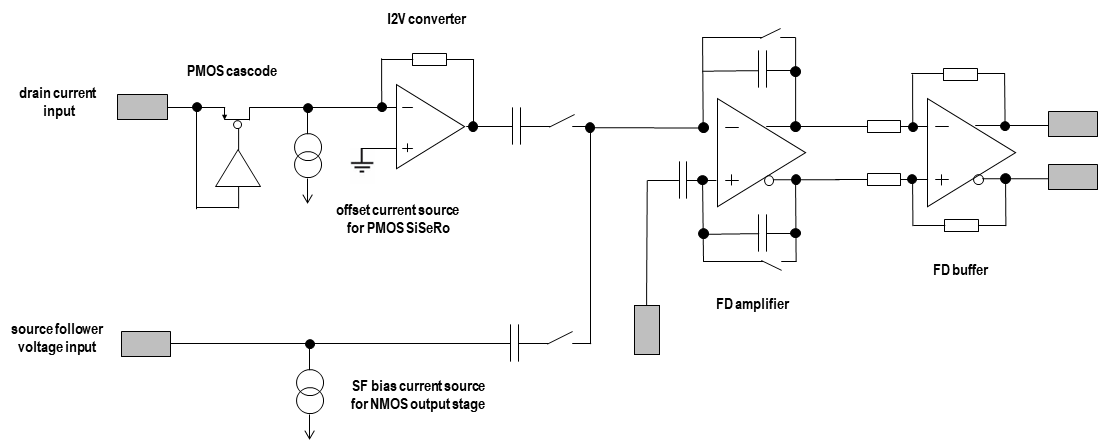}
    \caption{}
    \label{fig:mcrcsch}
    \end{subfigure}
  \begin{subfigure}{.35\textwidth}
     \includegraphics[width=\linewidth]{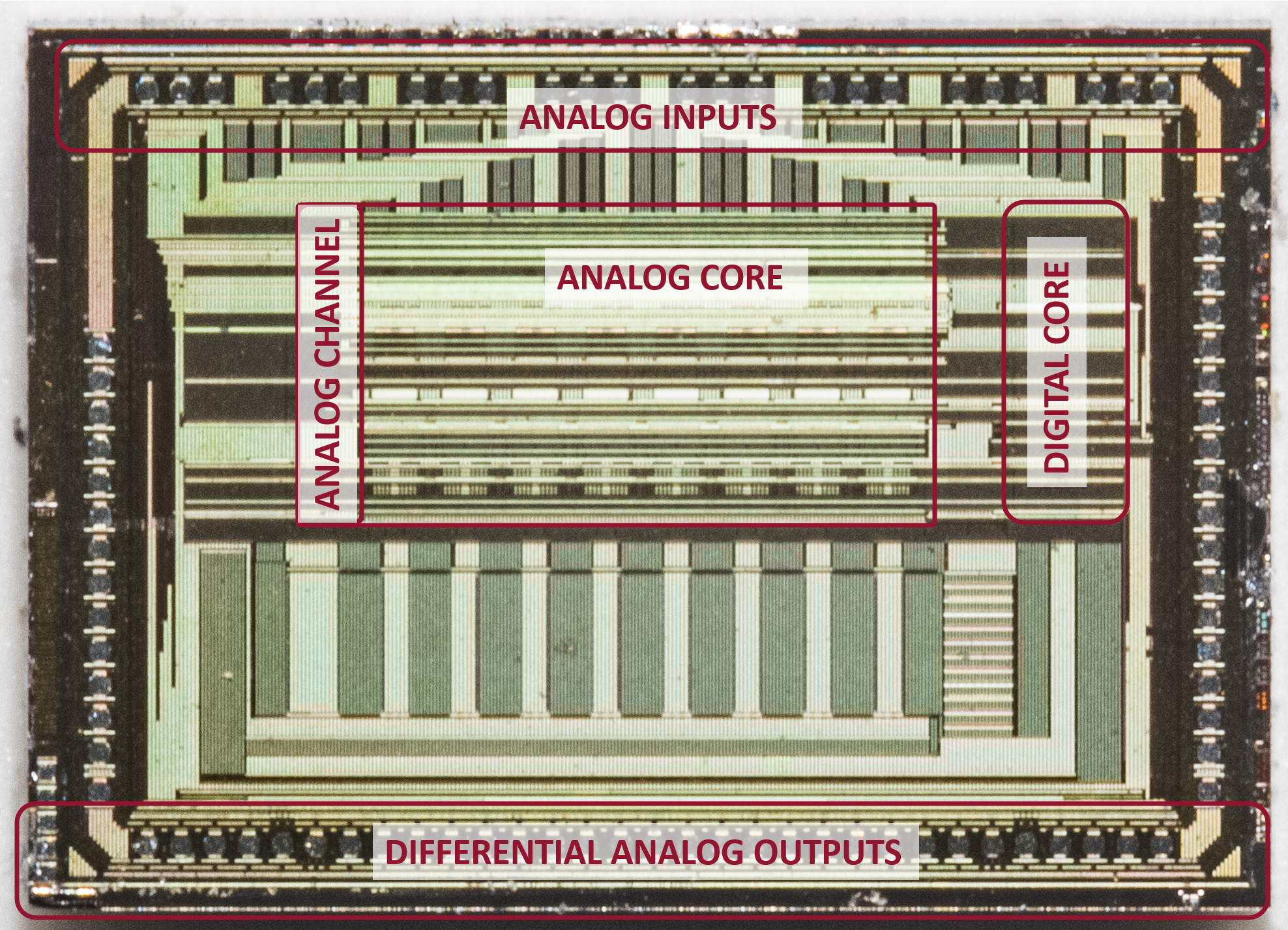}
    \caption{}
    \label{fig:mcrcphoto}
  \end{subfigure}%
    \caption{The MCRC-V1 single analog channel schematic (a) \cite{herrmann20_mcrc}, and (b) a photograph of the fabricated \textit{MCRC-V1} die with labels showing sub-circuit locations. Each of the analog channels starts with an input stage, which allows the user to select between the voltage and the current inputs, respectively. This input stage is followed by a preamplifier and a fully differential buffer driver. Eight channels plus a digital control section constitute the full chip.}
    \label{fig:mcrcschandlay}
\end{figure}

The MCRC-V1 features 8 analog channels. Each channel has two inputs: a source follower voltage input (SF) and an experimental drain current (DR) input. Figure~\ref{fig:mcrcsch} shows a simplified schematic of a single analog channel. Both inputs have adjustable current sources to set the operating point (OP) of the corresponding detector, and are AC coupled to internal programmable switches, making it easy to shift between the two input circuits. Only one readout mode can be selected at a given time for the entire chip. The readout mode switches are followed by a fully differential preamplifier (PRE), which performs single-ended-to-differential conversion and provides gain. There are two user-selectable gain settings of 8 V/V (G8) and 16 V/V (G16), respectively. The output of the PRE is buffered to the chip output pins by a unity gain fully differential output buffer (OB) that can drive a 100 \unit{\ohm} impedance line with an almost rail-to-rail output swing, making it simple to interface to a commercial analog-to-digital (ADC) converter \cite{porelMCRCspie2022,herrmann20_mcrc}.           

Apart from the analog core, the MCRC-V1 also features the following:
\begin{itemize}
  \item A digital core with eight 16-bit data registers that form the chip's volatile memory. The registers can be programmed and/or read back through a serial peripheral interface (SPI) bus. The SPI interface is fully duplex and makes use of low-voltage differential signaling (LVDS) transceivers.  
  \item Four programmable 7-bit digital-to-analog (DAC)s to independently set each of the operating points of the active cascode (ACAS), current-to-coltage (I2V), PRE, and OB. All nodes are externally accessible to provide the option of precise calibration, monitoring or bypassing, if necessary.
  \item Test circuitry to access and monitor the most important nodes in the ASIC. This is used for debugging and characterization purposes, and is not intended to be used during normal operation.  
\end{itemize}

Figure~\ref{fig:mcrcphoto} shows a photograph of the fabricated MCRC-V1 die with dimensions of 4160 \unit{\um} by 2900 \unit{\um}. The chip has been designed and manufactured in the X-FAB ultra-low-noise CMOS XH035 3.3 V 350 \unit{\nano m} process technology with isolated p-wells \cite{xfab}. This makes it possible to keep the chip substrate isolated from the bulk of the n-type transistors, i.e. switching currents are not flowing through the substrate, preventing transistor transient disturbances from coupling to sensitive nodes of the ASIC and corrupting the signal of interest. Furthermore, all analog circuitry is surrounded by deep p-well guard rings that ensure good isolation and keep the bulk substrate potential stable to avoid return current ground loops. The input and output signal carrying wires are surrounded by metal connected to the substrate potential to prevent electromagnetic fields, created by the rapidly switching signals, from coupling between neighbouring channels. To avoid further noise contamination from the digital circuitry, the analog and digital voltage rails are kept separate \cite{porelMCRCspie2022}.  

\section{Radiation tolerance}
\label{sec:mcrc_and_rad00}

Radiation testing has been conducted at Brookhaven National Laboratory (BNL). Our measurement setup consisted of a Cobalt-60 source, an ASIC test motherboard, four printed circuit boards (PCB)s with MCRC-V1 ASICs installed, a power supply, and several blocks of lead for shielding. 

\begin{figure}
    \centering
    \begin{subfigure}{.437\textwidth}
    \includegraphics[width=\linewidth]{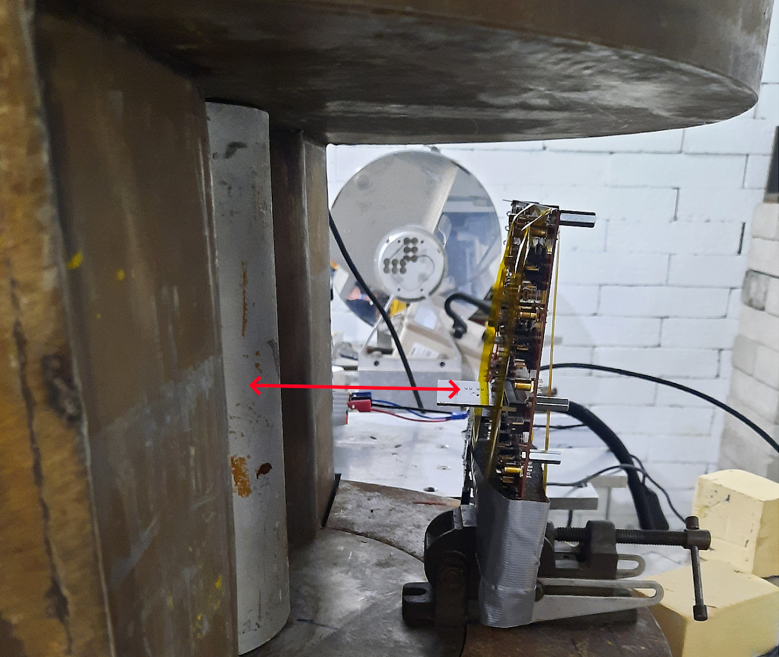}
    \caption{}
    \label{fig:mcrc_rad_side}
    \end{subfigure}
  \begin{subfigure}{.5\textwidth}
     \includegraphics[width=\linewidth]{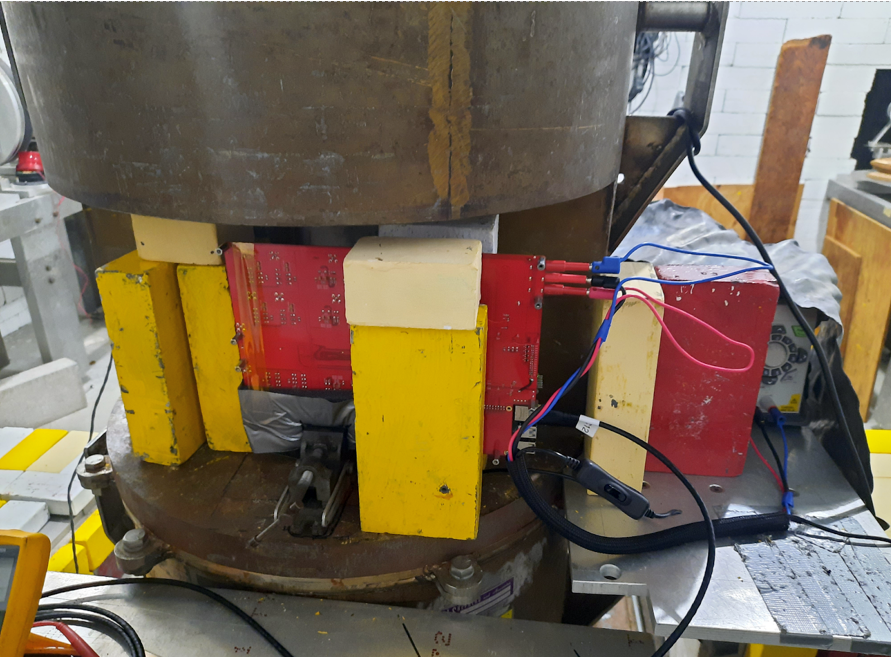}
    \caption{}
    \label{fig:mcrc_rad_behind}
  \end{subfigure}%
    \caption{The MCRC radiation test setup inside the hatch. (a) Side view without the lead shielding blocks that shows the alignment of the ASIC to the source position. (b) View from behind the setup with lead shielding blocks in place.}
    \label{fig:mcrc_rad_setup}
\end{figure}

The Cobalt-60 source emits gamma photons at energies of 1.19 MeV and 1.31 MeV, respectively. Dosimetry performed by facility staff at a distance of 4.25 inches from the source, estimated a dose rate of 9114 rad/h for the day of July 24th, 2023. To leverage the same dose, the ASIC has been carefully positioned in the same spot. Figures \ref{fig:mcrc_rad_side} and \ref{fig:mcrc_rad_behind} show the setup and the alignment of the chip to the source. Lead blocks were placed around the ASIC to shield the motherboard electronics. For safety purposes, the Cobalt-60 source is stored underground when people are working inside the hatch. The setup was operated by a computer sitting outside the hatch. The two were connected by a 50 feet long Ethernet cable. After the setup passed the safety and operational check-list, the hatch was sealed with an interlock system. Upon interlock system activation, the source was raised and the tests were executed sequentially for each of the four ASIC boards. Each MCRC-V1 chip was randomly selected from the original lot and directly wirebonded to its board. The exposure periods for each board were approximately 5.5 hours long, corresponding to a total ionizing dose of roughly 50 krad. The chips were powered on and fully configured for the duration of the tests. 

To identify, characterize, and time stamp a failure mechanism, certain ASIC parameters were contentiously monitored and logged on the motherboard. Specifically, these included the supply currents from all voltage rails (VDDA, VDDB and VDDD), the operating points (bias voltages) of all four main analog signal chain sub-circuits (the active cascode, the I2V amplifier, the preamplifier, and the output buffer). In addition, some internal registers of the MCRC-V1 were monitored for bit errors by continuously writing and reading back random digital data. We note that during the testing of the first chip, one of the two diagnostic ADCs that were part of the test motherboard failed after approximately 2.5 hours of exposure time. It was later determined that this failure occurred due to the position of that component being close to the ASIC under test, where the shielding was insufficient. Subsequently, three out of four biases were rewired to a set of unused auxiliary inputs on the second ADC.   

\begin{figure}[ht]
\begin{center}
\begin{tabular}{c} 
\includegraphics[width=\linewidth]{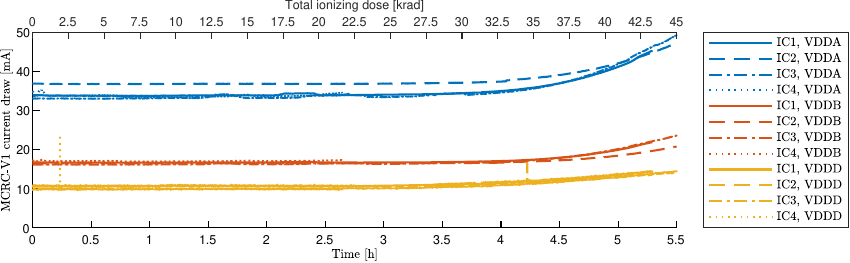}
\end{tabular}
\end{center}
\caption{Currents of the MCRC-V1 supply voltage rails as functions of time and TID for the four MCRC v1 chips tested. The current consumption stays almost constant up to radiation doses of $\sim$ 25krad. } 
\label{fig:mcrc_rad_curr}
\end{figure}

\begin{figure}[ht]
\begin{center}
\begin{tabular}{c} 
\includegraphics[width=\linewidth]{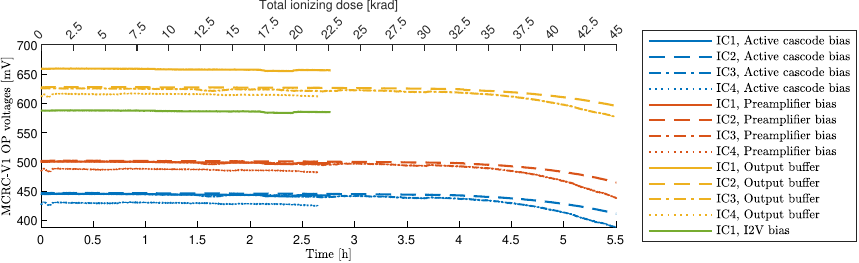}
\end{tabular}
\end{center}
\caption{MCRC-V1 analog circuitry bias voltages as functions of time and TID for the four chips tested. The bias voltages are a proxy for threshold voltage shifts of the bias transistors. No significant change is measured for radiation doses up to $\sim$ 25 krad, after which  voltages deviate by up to $\sim$ 50 mV for 45 krad.}  
\label{fig:mcrc_rad_bias}
\end{figure}

All four chips survived the radiation exposure. No bit errors were detected on the digital interface. Figures \ref{fig:mcrc_rad_curr} and \ref{fig:mcrc_rad_bias} show the supply currents and analog circuitry biases as functions of time and TID for all four chips. The chip circuitry remained unaffected until a TID dose of approximately 25 krad had been reached. Following higher doses, a shift in the bias voltages and supply currents was observed.

\begin{figure}[ht]
    \centering
    \begin{subfigure}{.49\textwidth}
    \includegraphics[width=\linewidth]{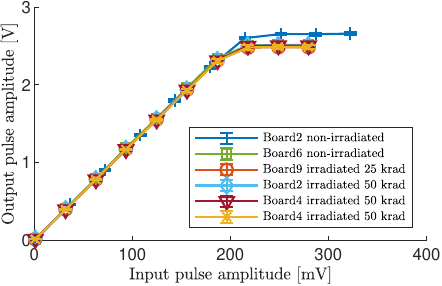}
    \caption{}
    \label{fig:mcrc_rad_p2}
    \end{subfigure}
  \begin{subfigure}{.45\textwidth}
     \includegraphics[width=\linewidth]{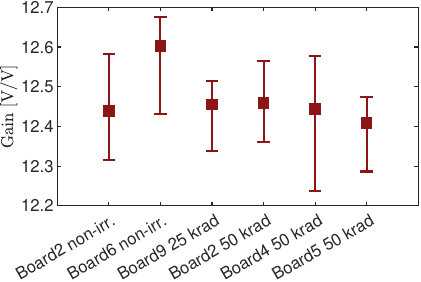}
    \caption{}
    \label{fig:mcrc_rad_p3}
  \end{subfigure}%
    \caption{Performance tests before and after irradiation. (a) Output pulse amplitude as a function of input pulse amplitude, showing the linearity of the MCRC V1 ASIC response. (b) Comparison of gain variation between irradiated and non-irradiated chips.}
    \label{fig:mcrc_rad_setup}
\end{figure}

To further quantify the impact of the radiation doses on performance, measurements of the MCRC-V1 analog chain were performed before and after radiation exposure for all chips under test. Figure \ref{fig:mcrc_rad_p2} shows the output pulse amplitude as a function of input pulse amplitude. Board 2 indicates a lowered saturation threshold between the non-irradiated and irradiated pulse response, although the difference is very small. Figure \ref{fig:mcrc_rad_p3} shows the measured gain for all four irradiated boards along with two non-irradiated chips. The gain is very uniform with minimal differences between the irradiated and non-irradiated boards. That is, the observed differences between irradiated and non-irradiated devices falls within the channel-to-channel variation that is inherent to the manufacturing process. We conclude that no noticeable performance degradation between irradiated and non-irradiated chips was observed for TID doses of up to 50 krad.         

\FloatBarrier
\section{Operational experience and test results}
\label{sec:workingwithmcrc}

The baseline performance of the standalone MCRC-V1 readout ASIC has been measured and reported in \citenum{porelMCRCspie2022}. A summary of the results is given in Table \ref{tab:MCRCsummary}.

\begin{table}[ht]
\caption{Summary of the MCRC-V1 performance parameters.} 
\label{tab:MCRCsummary}
\begin{center}       
\begin{tabular}{|l||c|c|} 
\hline
\rule[-1ex]{0pt}{3.5ex}  \textbf{Parameter} & \multicolumn{2}{|c|}{\rule[-1ex]{0pt}{3.5ex} \textbf{Measured Value}} \\
\hline \hline
\rule[-1ex]{0pt}{3.5ex}  \textbf{Gain} & \multicolumn{2}{|c|}{\rule[-1ex]{0pt}{3.5ex} $<$G16$>$ = 12.14 $\pm$ 0.35 V/V} \\
\rule[-1ex]{0pt}{3.5ex}                & \multicolumn{2}{|c|}{\rule[-1ex]{0pt}{3.5ex} $<$G8$>$ = 6.17 $\pm$ 0.59 V/V} \\
\hline
\rule[-1ex]{0pt}{3.5ex}  \textbf{Input dynamic range} & \multicolumn{2}{|c|}{\rule[-1ex]{0pt}{3.5ex} SF input: $\approx$ 320 \unit{\milli V_{PP}}} \\
\rule[-1ex]{0pt}{3.5ex}   & \multicolumn{2}{|c|}{\rule[-1ex]{0pt}{3.5ex} DR input: $\approx$ 35 \unit{\micro A_{PP}} ($\approx$ 500 \unit{\micro A_{PP}} with DRCS offset)} \\
\hline
\rule[-1ex]{0pt}{3.5ex}  \textbf{Input referred noise} & \multicolumn{2}{|c|}{\rule[-1ex]{0pt}{3.5ex} $<$\unit{N_{RMS}}$>$ = 56.94 $\pm$ 13.47 \unit{\micro V_{RMS}}} \\
\rule[-1ex]{0pt}{3.5ex}  & \multicolumn{2}{|c|}{\rule[-1ex]{0pt}{3.5ex} $<$\unit{N_{RMS}}$>$ = 1.63 $\pm$ 0.38 \unit{e^{-}_{RMS}} @ k = 35 \unit{\micro V/e^{-}}} \\
\rule[-1ex]{0pt}{3.5ex}  & \multicolumn{2}{|c|}{\rule[-1ex]{0pt}{3.5ex} $<$\unit{N_{VSD}}$>$ = 6.18 $\pm$ 1.46 \unit{\nano V/\sqrt{Hz}}} \\
\hline
\rule[-1ex]{0pt}{3.5ex}  \textbf{Crosstalk} & \multicolumn{2}{|c|}{\rule[-1ex]{0pt}{3.5ex} $\leq$ -75 dB in passband} \\
\hline
\rule[-1ex]{0pt}{3.5ex}   & \unit{I_{OB\_BIAS}} = 75 \unit{\micro A} & \unit{I_{OB\_BIAS}} = 150 \unit{\micro A} \\
\hline

\rule[-1ex]{0pt}{3.5ex}  \textbf{Bandwidth} & G16: 42.75 $\pm$ 5.91 & G16: 86.05 \\\rule[-1ex]{0pt}{3.5ex}  & G8: 54.95 $\pm$ 7.30 & G8: 116.38 \\
\hline

\rule[-1ex]{0pt}{3.5ex}  \textbf{Power consumption} & 30.8 mW/ch & 39.6 mW/ch \\\cline{2-3}
\rule[-1ex]{0pt}{3.5ex}   & \multicolumn{2}{|c|}{\rule[-1ex]{0pt}{3.5ex} Digital core: 37.2 mW}  \\

\hline
\end{tabular}
\end{center}
\end{table} 

To develop the user case of the MCRC as a high-speed, low-noise, and low-power readout-on-a-chip system, the ASIC prototype has been simultaneously deployed as a replacement for discrete electronics at both Stanford and MIT. In the Stanford case, it has been paired with single channel CCID93 detectors, which feature one p-jfet output similar in construction to the outputs on the CCID85 detector \cite{10.1117/1.JATIS.8.2.026005}. The primary purpose was  to make direct performance-wise comparisons of the ASIC to its discrete electronics counterparts and to optimize its operating points. The MIT tests had the MCRC-V1 paired with an 8 channel CCID89 detector to evaluate and gain confidence in its multichannel operation.

\subsection{Single channel operation and optimization}
\label{sec:mcrc_and_ccid93}

\begin{figure}
    \centering
    \begin{subfigure}{.2\textwidth}
        \includegraphics[width=\linewidth]{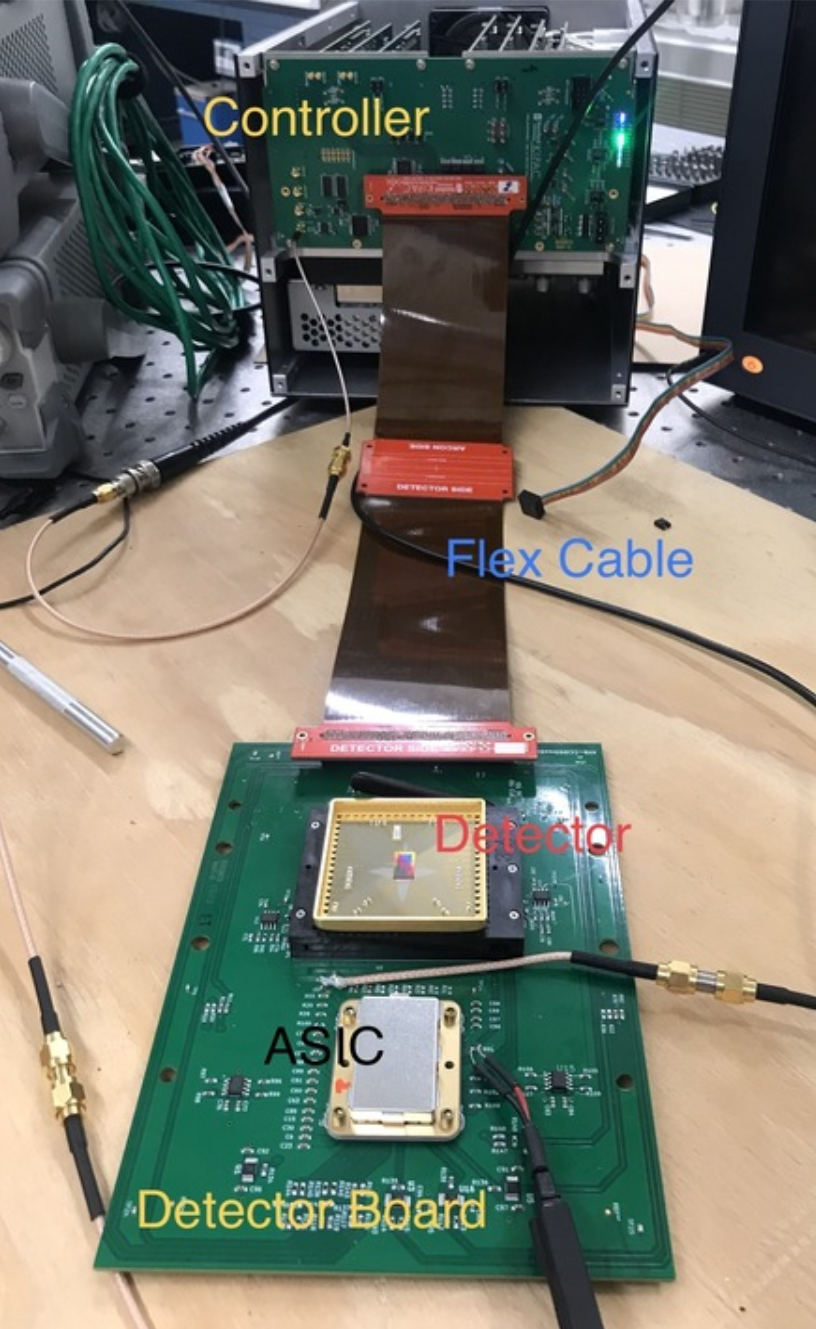}
        \caption{}
        \label{fig:beamline1}
        \end{subfigure}
    \begin{subfigure}{.28\textwidth}
        \includegraphics[width=\linewidth]{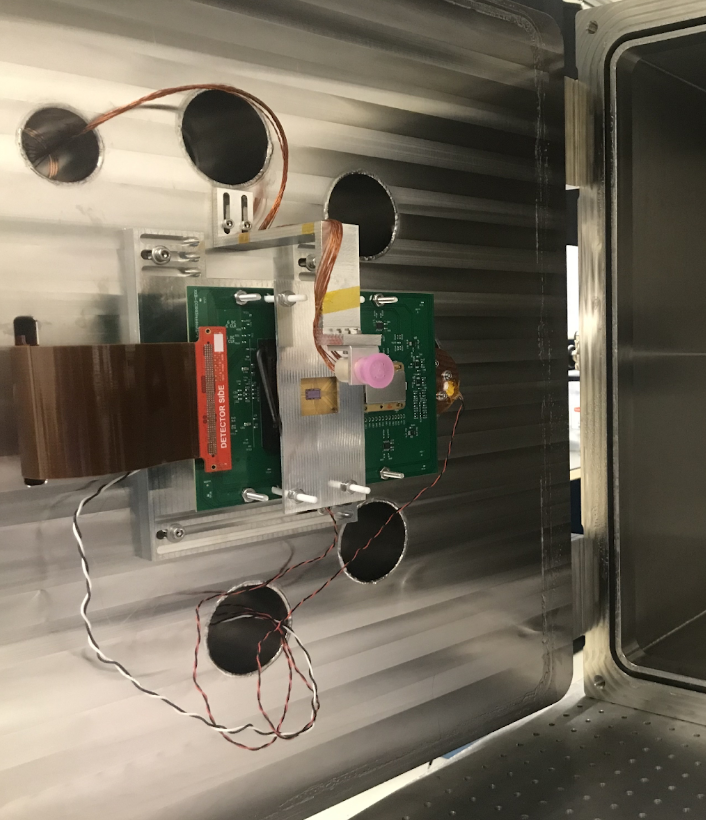}
        \caption{}
        \label{fig:beamline2}
        \end{subfigure}%
    \begin{subfigure}{.435\textwidth}
        \includegraphics[width=\linewidth]{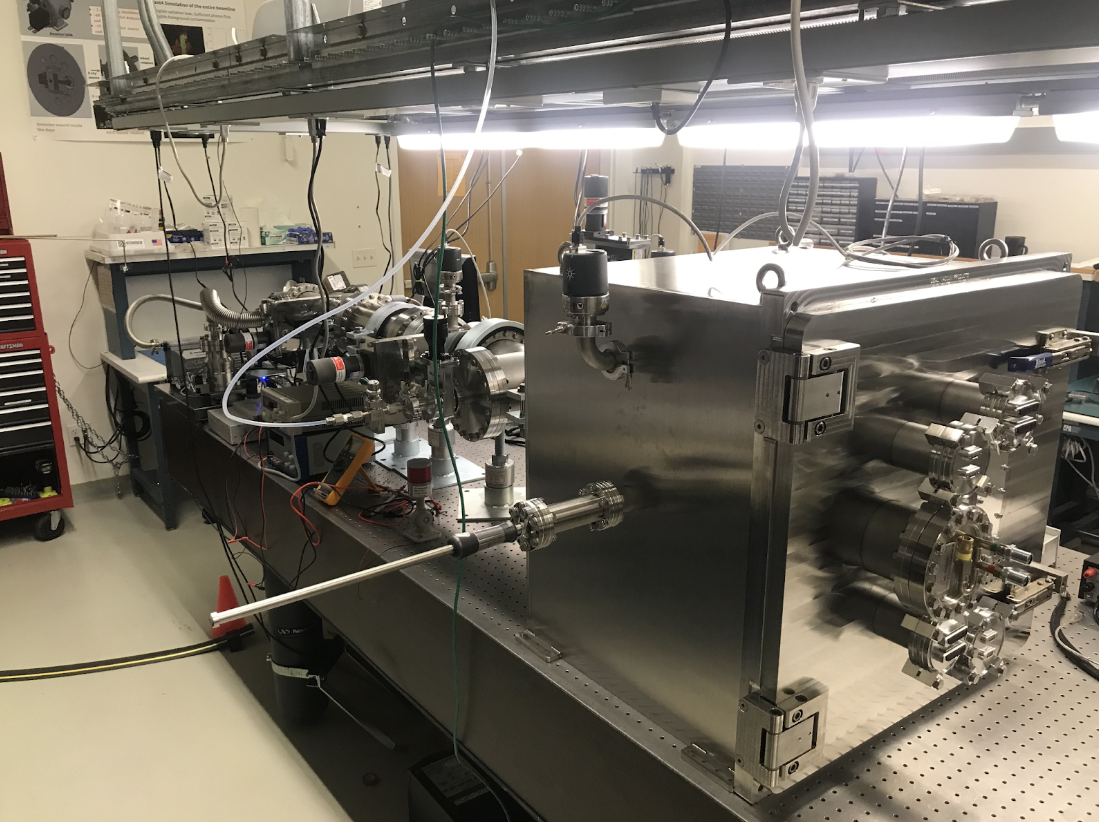}
        \caption{}
        \label{fig:beamline3}
        \end{subfigure}%
    \caption{The test setup with electronic units connected to form the readout system. (a) A custom flex lead connects the detector board inside (b) the chamber to the outside. The beamline (c) is designed to efficiently produce monoenergetic X-ray fluorescence lines in the 0.3 - 10 keV energy range, while maintaining detector temperatures as low as 173 K \cite{HStueberetal2024}.}
    \label{fig:beamline_setup}
\end{figure}

The characterization and optimization tests were carried out in the newly developed and commissioned 2.5 meter-long X-ray beamline test system \cite{HStueberetal2024}. An STA Archon CCD controller was used, which features 8 ADC channels and a variety of voltage biasing options and clock lines to control and drive the CCD \cite{STAArchon}. Figure \ref{fig:beamline1} shows the electronics part of the test setup. The CCID93 detector is mounted on the detector board by means of a zero insertion force (ZIF) socket connector. Similarly, to achieve maximum modularity, the MCRC-V1 has been wirebonded to a daughter PCB board which is mounted on the detector board without soldering. The connection is achieved using a Samtec high-density 1.27 mm thick interposer \cite{GMIsamtec}, that provides the lowest parasitic capacitance possible on a modular interface. The detector board acts solely as an interconnect layer between the detector, MCRC-V1, and the ADCs in the Archon controller. To that end, a rigid-flex PCB board is used to connect the detector board from inside the vacuum to the STA Archon controller outside, as shown in Figure \ref{fig:beamline2}. Figure \ref{fig:beamline3} shows a picture of the entire beamline from the X-ray source on the left to the chamber door on the right.     

System performance parameters such as gain, read noise, and full-width-at-half-maximum (FWHM) energy resolution were determined at X-ray line energies of 1.5 keV, 5.9 keV, 8.04 keV and 8.9 keV, using Aluminium (Al), Iron-55 ($^{55}$Fe) and Copper (Cu) sources. In addition, a Teflon source was used to calculate the maximum observed continuum energy (MOCE) as reported in row 7 of Table \ref{tab:gainnoisereslut_table}. The low gain setting (G8) of the MCRC-V1 has been measured to be approximately 6 V/V, while the high gain setting (G16) is 12 V/V, which matches the standalone measurements summarized in Table \ref{tab:MCRCsummary}. These gains are lower than the nominal design values of 8 V/V and 16 V/V, respectively, with the differences traced to a voltage divider effect caused by the output series resistance of the MCRC-V1 output buffer and the 100 \unit{\ohm} parallel termination.

\begin{table}
\caption{Test results from system response optimization.} 
    \centering
    \begin{tabular}{|>{\raggedright\arraybackslash}p{0.2\linewidth}||>{\centering\arraybackslash}p{0.1\linewidth}|>{\centering\arraybackslash}p{0.1\linewidth}|>{\centering\arraybackslash}p{0.1\linewidth}|>{\centering\arraybackslash}p{0.1\linewidth}|>{\centering\arraybackslash}p{0.1\linewidth}|>{\centering\arraybackslash}p{0.1\linewidth}|} \hline 
         \textbf{MCRC-V1 gain setting}&  \textbf{low}&  \textbf{low}&  \textbf{low}&  \textbf{high}&  \textbf{high}&  \textbf{high}\\ \hline 
         \textbf{Clamping voltage [V]}&  \textbf{+0}&  \textbf{+0.7}&  \textbf{+2}&  \textbf{+0}&  \textbf{+0.7}&  \textbf{+2}\\ \hline \hline 
         \textbf{Spectral peak \newline@1.5 keV [mV]}&  84.3&  84.4&  84.4&  166.5&  166.9&  167.0\\ \hline
         \textbf{Spectral peak \newline@5.9 keV [mV]}&  335.6&  335.6&  335.8&  658.2&  659.1&  658.7\\ \hline 
         \textbf{Spectral peak \newline@8.04 keV [mV]}&  457.8&  458.1&  457.9&  869.7&  869.1&  868.9\\ \hline 
         \textbf{System gain [\unit{\micro V}/\unit{e^{-}}]}&  \multicolumn{3}{|c|}{207.0 $\pm$ 1.39 } &  \multicolumn{3}{|c|}{402.7 $\pm$ 6.65}\\ \hline
 \textbf{MOCE [keV]}& 9.58& 9.64 & 9.57 & 9.54 & 9.75 &9.61 \\\hline
 \textbf{Dynamic range [keV]}& 27.9& 38.8& 59.1& 11.1& 18.1&27.0\\\hline
 \textbf{Noise [$e^{-}$]}& 3.67 ($^{55}$Fe)
3.46 (Cu) 
3.37 (Al)& 3.47 ($^{55}$Fe)
3.49 (Cu)
3.40 (Al)& 3.46 ($^{55}$Fe) 3.50 (Cu)
3.38 (Al)& 3.68 ($^{55}$Fe)
3.76 (Cu)
3.24 (Al)& 3.21 ($^{55}$Fe)
3.35 (Cu)
3.19 (Al)&3.21 ($^{55}$Fe)
3.48 (Cu) 
3.14 (Al)\\\hline
 \textbf{FWHM @1.5 keV [eV]}& 75.72& 75.05
& 74.16
& 74.58& 73.46&74.06\\\hline
 \textbf{FWHM @5.9 keV [eV]}& 122.81& 121.27& 123.27& 125.90& 111.49&117.48\\\hline
 \textbf{FWHM @8.04 keV [eV]}& 151.89& 145.45& 146.74& 160.26& 144.64&146.05\\\hline
    \end{tabular}
    \label{tab:gainnoisereslut_table}
\end{table}

\subsubsection{System response optimization}
As expected, the high gain setting approximately doubles the system gain, while adjusting the clamping voltage of the ADC input does not have a significant effect. The systematic uncertainty on the system gain is (very) small, confirming a (very) linear response over the energy range studied. At high gain, the higher signal to noise ratio (SNR) delivers an overall lower readout noise. The dynamic range of the system is limited by the range of the 16-bit Archon ADCs. That is, the full scale input voltage range is constrained at the upper limit of 4 \unit{\volt _{PP}} while the lower limit is set by the distance between the bottom of the ADC range and the input signal baseline. The system dynamic range can be increased by decreasing the gain or increasing the clamping voltage, which shifts the baseline from the center towards one of the rails, depending on the input signal polarity. We note that during actual operations, the MCRC-V1 may have been saturated before the maximum range was reached for the larger range spans. The high gain setting with a clamping voltage of +0 V provides an energy spectrum range of up to 13 keV. The power draw of the system was unaffected by the gain or clamping voltage settings studied.

Operating point optimisation was carried out to determine the best ASIC performance values in terms of gain, noise, speed, and current draw. The MCRC-V1 analog circuitry has three adjustable parameters that can affect not only the ASIC but also the detector output response. More precisely, the ASIC has an input current source (SFCS), for which the source voltage (\unit{V_{SFCS\_SOURCE}}) can be externally adjusted to set the proper operating point of the detector output. The other two adjustable parameters are: the preamplifier bias current (\unit{I_{PRE\_BIAS}}) and the output buffer bias current (\unit{I_{OB\_BIAS}}). Unlike the \unit{V_{SFCS\_SOURCE}}, these two can be set by internal DACs. To optimize the overall system response, all three parameters were varied and the corresponding performance analyzed. 
Gain and noise performances were determined by taking 200 illuminated frames followed by 50 dark frames. Speed was evaluated by comparing the raw transient waveform settling behaviour at each bias point. For the transient figure, in general, a smaller reset peak indicates that the system is slew-limited (large signal response), while a shallower transition between the baseline and signal means the system is bandwidth-limited (small signal response).

\begin{figure}
    \centering
    \begin{subfigure}{.49\textwidth}
    \includegraphics[width=\linewidth]{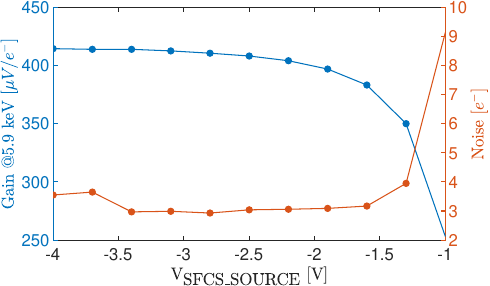}
    \caption{}
    \label{fig:mcrcopt_ncurv1}
    \end{subfigure}
  \begin{subfigure}{.405\textwidth}
     \includegraphics[width=\linewidth]{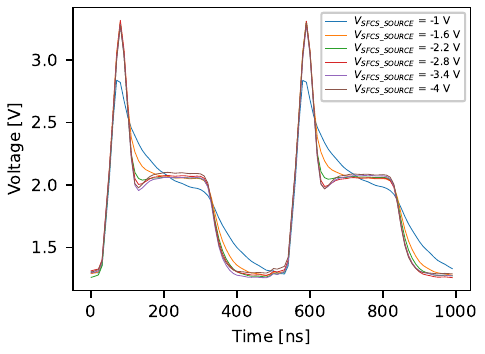}
    \caption{}
    \label{fig:mcrcopt_ncurv2}
  \end{subfigure}%
    \caption{(a) Gain and noise as functions of SFCS source voltage (NCUR\_V pin). (b) Transient waveform for \unit{V_{SFCS\_SOURCE}} voltages ranging from -1 V to -4 V. The MCRC-V1 operating point was: gain = 12 V/V, \unit{I_{PRE\_BIAS}} = 75 \unit{\micro A}, \unit{I_{OB\_BIAS}} = 50 \unit{\micro A}.}
    \label{fig:mcrcopt_ncurv}
\end{figure}

Figure \ref{fig:mcrcopt_ncurv1} shows the gain and noise of the system as functions of the SFCS source voltage applied to the MCRC-V1 pin named NCUR\_V. Gain performance improves as the \unit{V_{SFCS\_SOURCE}} increases, while the optimal noise performance is achieved for \unit{V_{SFCS\_SOURCE}} voltages between -3 V and -2 V. Figure \ref{fig:mcrcopt_ncurv2} shows a set of transient waveforms for various \unit{V_{SFCS\_SOURCE}} voltages. A clear effect on both the slew rate and bandwidth is observed, where a maximum \unit{V_{SFCS\_SOURCE}} voltage of -2 V is needed to maintain a fast enough response. The obtained optimal \unit{V_{SFCS\_SOURCE}} value is -2.8 V.  

\begin{figure}
    \centering
    \begin{subfigure}{.49\textwidth}
    \includegraphics[width=\linewidth]{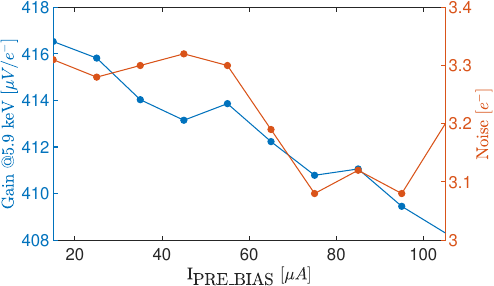}
    \caption{}
    \label{fig:mcrcopt_ipre1}
    \end{subfigure}
  \begin{subfigure}{.38\textwidth}
     \includegraphics[width=\linewidth]{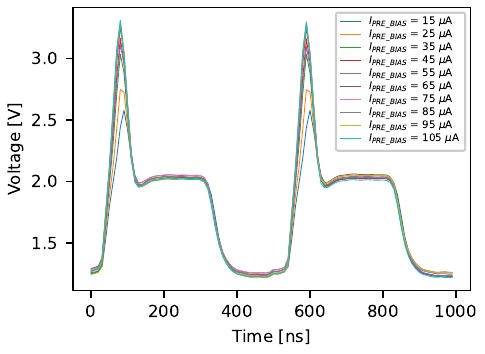}
    \caption{}
    \label{fig:mcrcopt_ipre2}
  \end{subfigure}%
    \caption{(a) Gain and noise as functions of the preamplifier bias current. (b) Transient waveform for \unit{I_{PRE\_BIAS}} ranging from 15 \unit{\micro A} V to 105 \unit{\micro A}. The MCRC-V1 operating point was: gain = 12 V/V, \unit{V_{SFCS\_SOURCE}} = -2.8 V, \unit{I_{OB\_BIAS}} = 50 \unit{\micro A}.}
    \label{fig:mcrcopt_ipre}
\end{figure}

Figure \ref{fig:mcrcopt_ipre1} shows gain and noise as functions of the preamplifier bias current. Overall, \unit{I_{PRE\_BIAS}} does not appear to have a significant affect on the performance. However, the gain exhibits a worsening trend with increasing bias current, while the noise seems to be improving. Figure \ref{fig:mcrcopt_ipre2} shows input signal transient waveforms for different preamplifier bias current settings. A minimum bias current of 35 \unit{\uA} is required to maintain the nominal slew rate of the system.

\begin{figure}
    \centering
    \begin{subfigure}{.49\textwidth}
    \includegraphics[width=\linewidth]{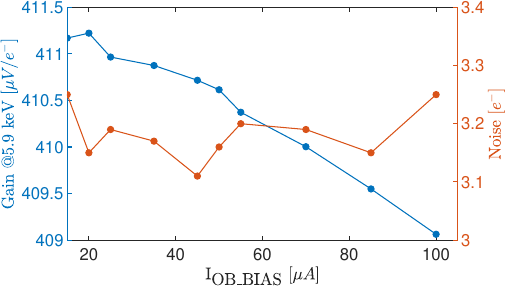}
    \caption{}
    \label{fig:mcrcopt_iob1}
    \end{subfigure}
  \begin{subfigure}{.38\textwidth}
     \includegraphics[width=\linewidth]{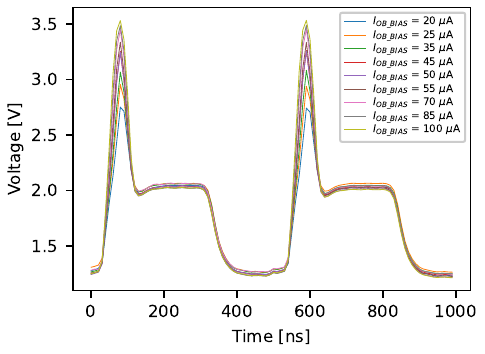}
    \caption{}
    \label{fig:mcrcopt_iob2}
  \end{subfigure}%
    \caption{(a) Gain and noise as functions of the output buffer bias current. (b) Transient waveforms for \unit{I_{OB\_BIAS}} ranging from 20 \unit{\micro A} V to 100 \unit{\micro A}. The MCRC-V1 operating point was: gain = 12 V/V, \unit{V_{SFCS\_SOURCE}} = -2.8 V, \unit{I_{PRE\_BIAS}} = 75 \unit{\micro A}.}
    \label{fig:mcrcopt_iob}
\end{figure}

\begin{figure}
\begin{center}
\begin{tabular}{c} 
\includegraphics[width=6.5in]{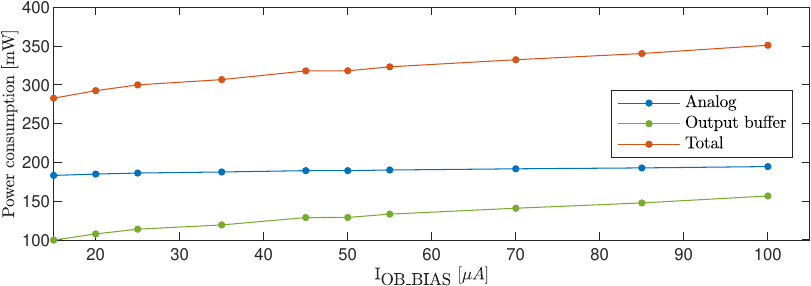}
\end{tabular}
\end{center}
\caption{MCRC-V1 power consumption for the analog (VDDA/VSSA) rail, output buffer (VDDB/VSSB), and the total ASIC power consumption. Digital power consumption is not displayed, remaining stable at $\sim$ 34 mW.}
\label{fig:mcrcopt_power}
\end{figure}

Figure \ref{fig:mcrcopt_iob1} shows the gain and noise as functions of the output buffer bias current. Both the gain and the noise do not appear to be affected by the change in \unit{I_{OB\_BIAS}}. However, it does affect the slew rate of the buffer, as can be seen in Figure \ref{fig:mcrcopt_iob2}. A minimum of 50 \unit{\micro A} is necessary to properly buffer the analog signal. In addition, an increasing \unit{I_{OB\_BIAS}} affects the power draw of the ASIC. Both the buffer rail (VDDB/VSSB) and analog circuitry rail (VDDA/VSSA) power consumption increases linearly with increasing output buffer bias current, as shown in Figure \ref{fig:mcrcopt_power}.

The MCRC-V1 preamplifier has a switched capacitor based feedback that needs to be periodically reset to baseline to avoid saturation. The reset can be applied at the end of each CCD line or at the end of every pixel window. Both cases show a comparable performance in terms of gain and noise. Similarly, sweeping the reset pulse width does not appear to have a significant effect. On the other hand, a minimum reset pulse width of 40 ns is necessary to provide an adequate transient response. The best performance, without sacrificing readout time, is achieved with a reset pulse width of 70 ns. We note that the ability to reset the readout ASIC at the pixel-by-pixel level might become necessary for large CCD detectors.

Table \ref{tab:gainnoisereslut_table} summarizes the results for gain, noise, dynamic range, MOCE, and FWHM.  

\subsubsection{Saturation recovery}
To ensure that minimal dead time occurs, the MCRC-V1 amplification chain requires prompt recovery from saturation events. On the detector side, the reset clock (RG) pulse drains the charge packet away from the output stage of the CCD after reading out each pixel. In normal operation, the RG pulse is too narrow (10 ns) for the system to saturate. However, increasing the RG reset pulse width and measuring the amplitude of the RG pulse directly has been found to saturate the MCRC-V1. More accurately, saturation becomes apparent for an RG pulse amplitude of $\sim$ 1.27 V, when the RG pulse width is greater than 100 ns. Nevertheless, the MCRC-V1 recovers with only marginal degradation in gain and noises, providing an overall performance comparable to that under the normal operation.  

\subsubsection{Overall system noise}
The effects of the MCRC-V1 on the overall system noise were also explored. Configuration wise, chip channel 6 was connected to the CCD p-jfet output, while five other channels had their inputs floating. The remaining two channels were disconnected from the Archon ADC, leaving the ADC inputs floating. To provide continuous time domain data without transient artifacts, the RG reset signal was kept high, meaning no actual charge packets were read out. A PSD analysis was then applied to obtain the noise spectral density of each channel. Table \ref{tab:mcrcnoisevschannel_table} reports the measurement results.

\begin{table}
\caption{Summary of noise measurements with the MCRC-V1 over all channels in various configurations.} 
    \centering
    \begin{tabular}{|>{\raggedright\arraybackslash}p{0.15\linewidth}||>{\centering\arraybackslash}p{0.22\linewidth}|>{\centering\arraybackslash}p{0.1\linewidth}|>{\centering\arraybackslash}p{0.2\linewidth}|>{\centering\arraybackslash}p{0.1\linewidth}|} \hline 
         \textbf{channel}&  \textbf{configuration} &  \textbf{IRN [nV]} &  \textbf{thermal noise floor [nV/$\sqrt{Hz}$]} &  \textbf{thermal noise [$e^{-}$]} \\ \hline \hline
         \textbf{CH1}& ASIC+ADC & 937.4 & 5.66 &  0.75 \\ \hline
         \textbf{CH2}& ADC only & 109.12 &  1.52 &  0.20 \\ \hline 
         \textbf{CH3}& ASIC+BUFF+ADC & 1104.9 & 5.97 & 0.79 \\ \hline 
         \textbf{CH4}& ASIC+BUFF+ADC & 1041.1 & 6.06 & 0.80 \\ \hline
         \textbf{CH5}& ASIC+ADC & 924.1 & 5.61 & 0.74 \\ \hline
         \textbf{CH6}& Detector+ASIC+ADC & 16070.6 & 12.92 & 1.71 \\\hline
         \textbf{CH7}& ASIC+ADC & 909.6 & 5.66 & 0.75 \\ \hline
         \textbf{CH8}& ADC only & 96.5 & 1.42 & 0.19 \\ \hline
         \textbf{CH1 - CH2}& ASIC only (derived)& 931.1 & 5.45 & 0.72 \\ \hline
         \textbf{CH6 - CH1}& Detector only (derived)& 16043.3 & 11.60 & 1.53 \\ \hline

    \end{tabular}
    \label{tab:mcrcnoisevschannel_table}
\end{table}

\begin{figure}
\begin{center}
\begin{tabular}{c} 
\includegraphics[width=6.5in]{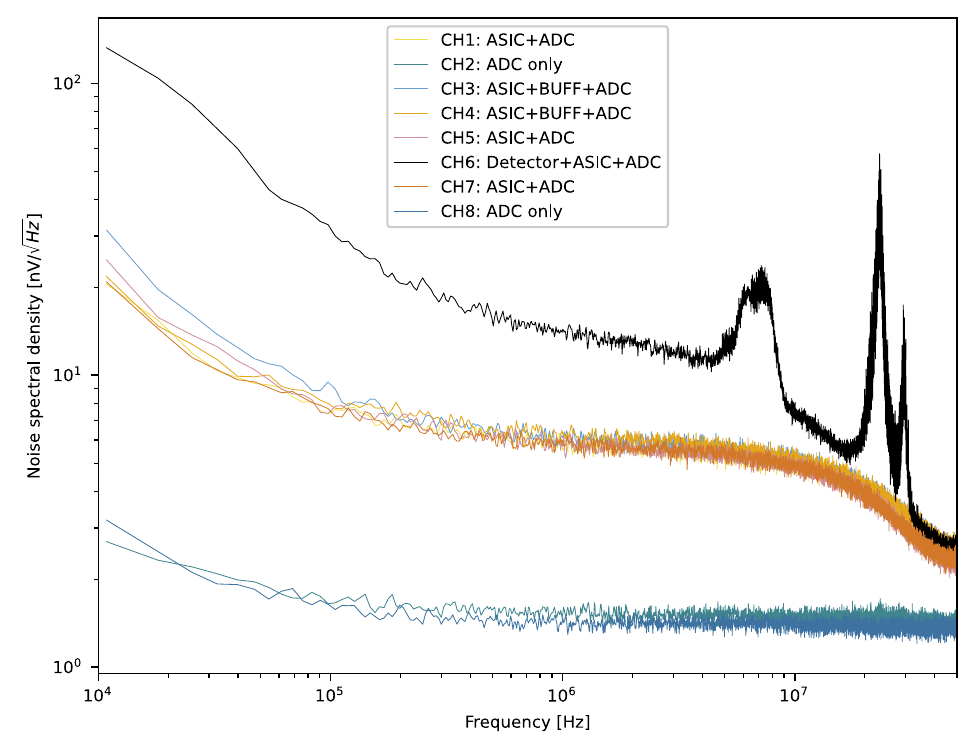}
\end{tabular}
\end{center}
\caption{Noise spectral density for all 8 channels of the MCRC-V1 in different configurations.}
\label{fig:mcrc_psd_8ch}
\end{figure}

Both the MCRC-V1 and Archon noise performance show a high level of consistency across all channels. Figure \ref{fig:mcrc_psd_8ch} shows the noise spectral density for all system channels. Channels 4 and 3 have a slightly higher noise spectral density due to the additional 100 \unit{\ohm} buffers. The ADC and MCRC-V1 noise spectra appear to be mostly thermal, while a larger flicker (1/f) noise contribution can be seen coming from the detector. The large bump at approximately 7 MHz originates from coupling with a non-grounded metal piece of the detector socket close to the detector output pin; once grounded the bump was amplitude reduced significantly. The components at higher frequencies are coupled to the system through the wire leads of the temperature sensors located on the detector board, as well as on the cold head of the detector mount.   

\subsubsection{Crosstalk isolation}
\begin{figure}
    \centering
    \begin{subfigure}{.49\textwidth}
    \includegraphics[width=\linewidth]{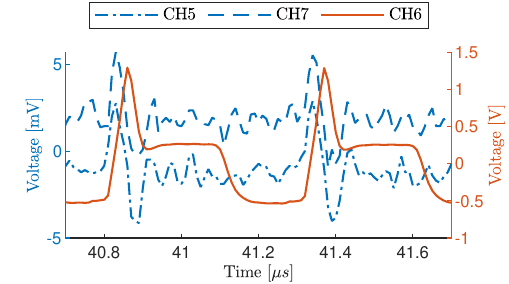}
    \caption{}
    \label{fig:mcrccrostalk1}
    \end{subfigure}
  \begin{subfigure}{.45\textwidth}
     \includegraphics[width=\linewidth]{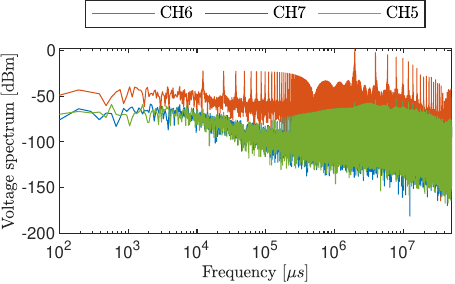}
    \caption{}
    \label{fig:mcrccrostalk2}
  \end{subfigure}%
    \caption{(a) Transient waveform showing the detector response on channel 6 and the neighbouring channels 5 and 7. (b) Voltage spectrum of channels 5, 6, and 7.}
    \label{fig:mcrccrostalk}
\end{figure}

\begin{figure}
    \centering
    \begin{subfigure}{.32\textwidth}
        \includegraphics[width=\linewidth]{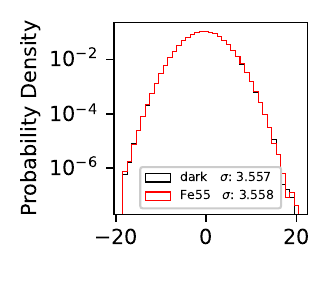}
        \caption{}
        \label{fig:hist1}
        \end{subfigure}
    \begin{subfigure}{.32\textwidth}
        \includegraphics[width=\linewidth]{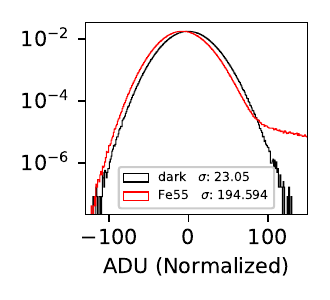}
        \caption{}
        \label{fig:hist2}
        \end{subfigure}%
    \begin{subfigure}{.32\textwidth}
        \includegraphics[width=\linewidth]{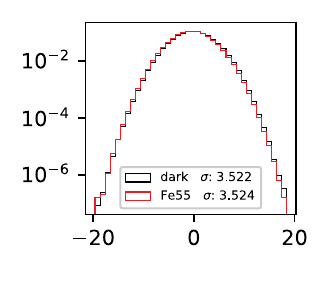}
        \caption{}
        \label{fig:hist3}
        \end{subfigure}%
    \caption{Pixel values in terms of probability density (dark and $^{55}$Fe frames) as functions of normalized analog digital units (ADU) or counts for the detector channel (b) and the adjacent unconnected MCRC-V1 channels 5 (a), and 7 (c), respectively.}
    \label{fig:mcrc_crisstalk_histFE55}
\end{figure}

Channel to channel crosstalk can significantly degrade system performance by introducing phantom "signals" in adjacent channels. Using the same channel configuration as in the noise tests, data were taken operating the detector and one ASIC channel operating in a normal mode (RG reset pulse nominal), while simultaneously gathering raw waveforms from the other channels. Dark, Iron-55 ($^{55}$Fe) and copper(Cu) frames were taken for each channel. Figures \ref{fig:mcrccrostalk1} and \ref{fig:mcrccrostalk2} show the transient waveforms and the corresponding spectra, respectively, for channels 5, 6, and 7. Channel 6 has the detector signal, while channels 5 and 7 are neighbouring channels. Evidence of capacitive crosstalk can be seen around the RG reset pulse in all of the channels connected, but the effect is most pronounced in channels 5 and 7. The measured isolation to the neighbouring channels is approximately 56 dBc, which is a lower value then the 75 dBc observed during the standalone MCRC-V1 measurements \cite{porelMCRCspie2022}. The exact coupling mechanism is currently unconfirmed. However, the board and detector layouts are the main suspects. Nevertheless, this crosstalk is only seen during pixel resets, where the acquired data is considered invalid. To investigate this further, images of the pixel values for each channel were examined. When the detector is dark or unconnected, the spread of the pixel values is Gaussian. Figures \ref{fig:mcrc_crisstalk_histFE55} a, b, and c show histogram-ed pixels for channels 5, 6, and 7, respectively. Comparing the dark and $^{55}$Fe source frame pixel values for unconnected channels, we see a negligible change in spread. The same has been observed for the Cu source. This confirms that crosstalk does not have a noticeable effect on the actual readout performance of the MCRC-V1, with 56 dB of channel-to-channel isolation representing an attenuation factor of $\sim$ 1000 V/V.

\subsection{Comparison with discrete electronics}

The detector board used is designed such that both the ASIC and the discrete preamplifier are available to the same detector output. The MCRC-V1 has been operated in the previously obtained optimal operating point: gain = 12 V/V, \unit{V_{SFCS\_SOURCE}} = -2.8 V, \unit{I_{PRE\_BIAS}} = 75 \unit{\micro A}, and \unit{I_{OB\_BIAS}} = 50 \unit{\micro A}. Table \ref{tab:mcrcvsdiscrete_table} reports the performance parameters (gain, noise, and FWHM) for each of the two cases cases.   

\begin{table}
\caption{Summary of gain, noise, and FWHM measurements with the MCRC-V1 and the discrete electronics.} 
    \centering
    \begin{tabular}{|>{\raggedright\arraybackslash}p{0.12\linewidth}||>{\centering\arraybackslash}p{0.1\linewidth}|>{\centering\arraybackslash}p{0.1\linewidth}|>{\centering\arraybackslash}p{0.1\linewidth}|>{\centering\arraybackslash}p{0.1\linewidth}|>{\centering\arraybackslash}p{0.05\linewidth}|>{\centering\arraybackslash}p{0.13\linewidth}|>{\centering\arraybackslash}p{0.1\linewidth}|}  \hline 
         \textbf{}&  \textbf{system gain @5.9 keV [\unit{\micro V/e^{-}}]} &  \textbf{electronics gain [V/V]} &  \textbf{detector conversion gain [\unit{\micro V/e^{-}}]} &  \textbf{FWHM @5.9 keV [eV]} &  \textbf{noise [$e^{-}$]} &  \textbf{power consumption [W]} &  \textbf{area footprint [\unit{\%}]} \\ \hline \hline
         \textbf{MCRC-V1}& 402.6 & 12 & 33.5 &  120.9 & 3.05 & 0.35 & 3.87 \\ \hline
         \textbf{Discrete}& 340.7 & 10 &  34.1 &  122.1 & 3.12 & 4 & 100 \\ \hline

    \end{tabular}
    \label{tab:mcrcvsdiscrete_table}
\end{table}

\begin{figure}[h]
    \centering
    \begin{subfigure}{.48\textwidth}
    \includegraphics[width=\linewidth]{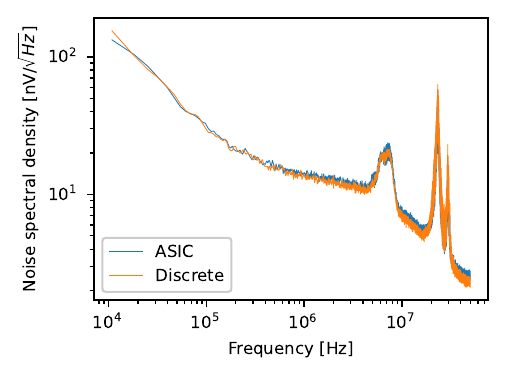}
    \caption{}
    \label{fig:mcrc_vs_discrete1}
    \end{subfigure}
  \begin{subfigure}{.49\textwidth}
     \includegraphics[width=\linewidth]{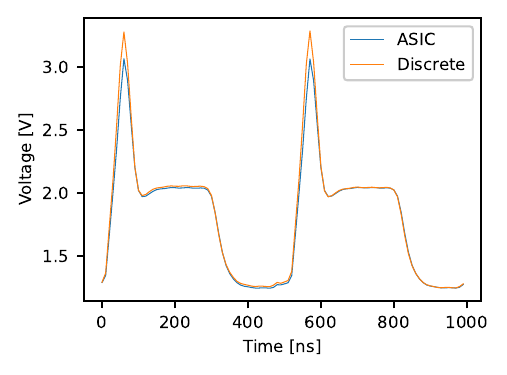}
    \caption{}
    \label{fig:mcrc_vs_discrete2}
  \end{subfigure}%
    \caption{Comparison plots of (a) Noise spectral density and (b) transient response of the MCRC-V1 and the discrete electronics implementation.}
    \label{fig:mcrc_vs_discrete}
\end{figure}

Figures \ref{fig:mcrc_vs_discrete1} and \ref{fig:mcrc_vs_discrete2} show the direct comparison of the MCRC-V1 readout ASIC and the discrete electronics implementation in terms of noise spectral density and speed, respectively. 

Our results confirm that the noise, gain, and speed performance of the MCRC-V1 is comparable to that of the discrete preamplifier. The total MCRC-V1 power consumption, at the optimal bias conditions, is $\sim$ 350 mW, while a single channel of the discrete preamplifier has a power consumption closer to 500 mW. Consequently, eight channels reach a power consumption of 4 W. The MCRC-V1 has an effective power consumption $\sim$ 11.5 times lower then the discrete preamplifier. In addition, the ASIC has a smaller footprint area than the layout of one single discrete preamplifier channel.

\subsection{Multichannel measurement results}
\label{sec:mcrc_CCID89}

\begin{figure}[h]
\begin{center}
\begin{tabular}{c} 
\includegraphics[width=6.5in]{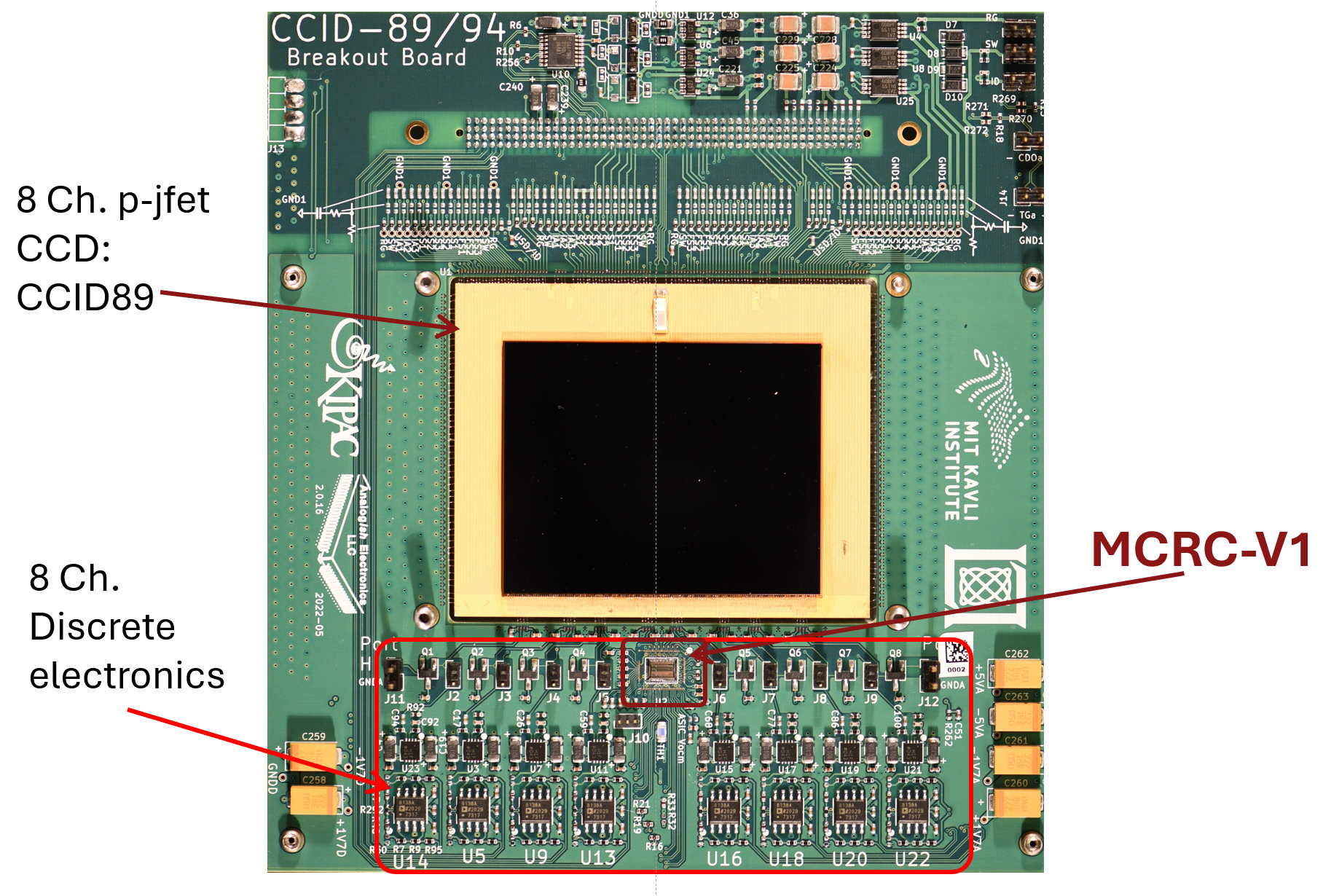}
\end{tabular}
\end{center} 
\caption{PCB board with 8 channel CCDID89 p-jfet output CCD with both the MCRC-V1 and discrete electronics.} 
\label{fig:mcrc_and_ccid89}
\end{figure}

The MCRC-V1 is an 8 channel readout ASIC. However, the tests described at Stanford employed only single output detectors. To test all 8 channels in parallel, an MCRC-V1 ASIC was also installed on an MIT-designed PCB board with an 8 output CCID89 CCD \cite{10.1117/1.JATIS.8.2.026005}. Figure \ref{fig:mcrc_and_ccid89} shows a photograph of the PCB board that hosts the CCID89 detector along with both the MCRC-V1 and discrete electronics implementation. 
Initial testing of the CCD board was carried out using the discrete component chains, When that was completed, the ASIC chip was installed and the discrete chains were disabled. Another round of testing was then performed. All 8 readout channels were found to work well, with performance quality very close to that demonstrated by with the discrete components, but at a greatly reduced power consumption. 

\begin{figure}[h]
    \centering
    \begin{subfigure}[t]{0.49\textwidth}
    \centering
        \includegraphics[width=3.4in]{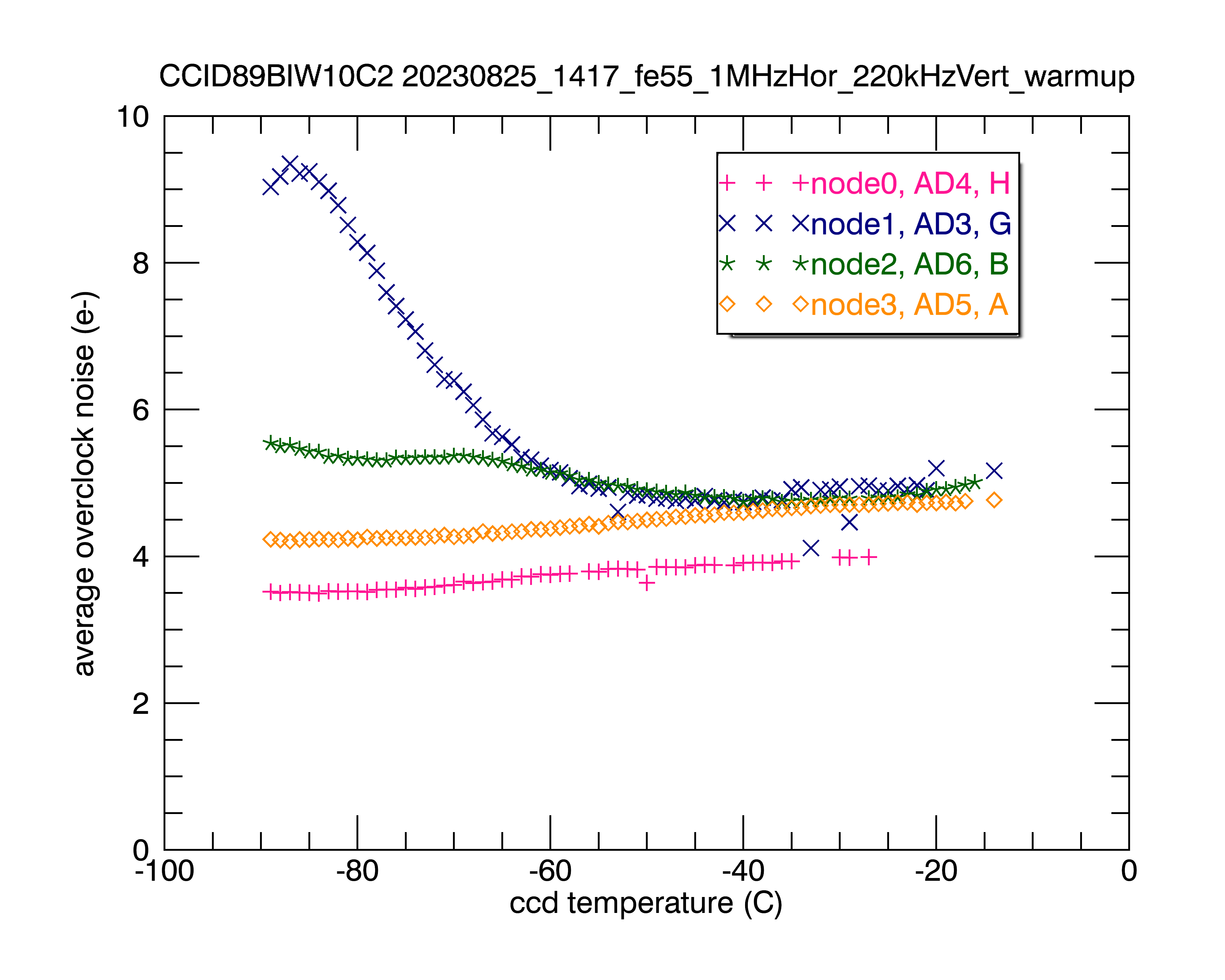}
    \caption{Discrete components readout}
    \label{}
    \end{subfigure}
    \begin{subfigure}[t]{0.49\textwidth}
    \centering
        \includegraphics[width=3.4in]{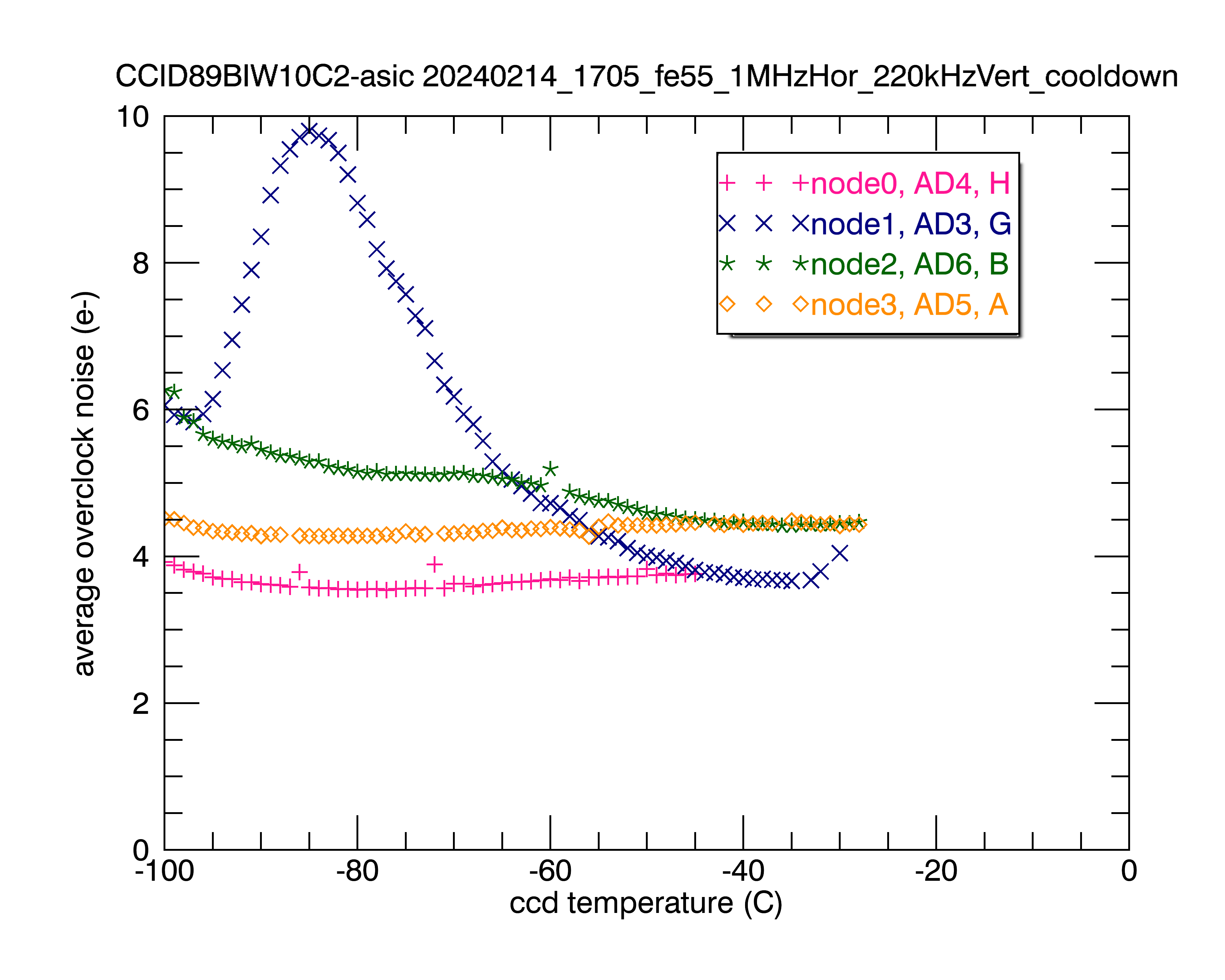}
    \caption{Readout with ASIC}
    \label{}
    \end{subfigure}
    \caption{Noise as a function of CCD temperature for 4 different output nodes.}
    \label{fig:ccid89_noise}
\end{figure}

CCD readout noise values were measured over a wide range of temperatures, both before and after ASIC installation for 4 CCD channels, as shown on Figure \ref{fig:ccid89_noise}. Noise units were converted into electrons using X-rays emitted by an $^{55}$Fe radioactive source. This conversion takes into account gain changes with temperature. The difference in noise between ASIC and discrete component readout is very small, a fraction of an electron.

\begin{figure}[h]
    \centering
    \includegraphics[width=1\textwidth]{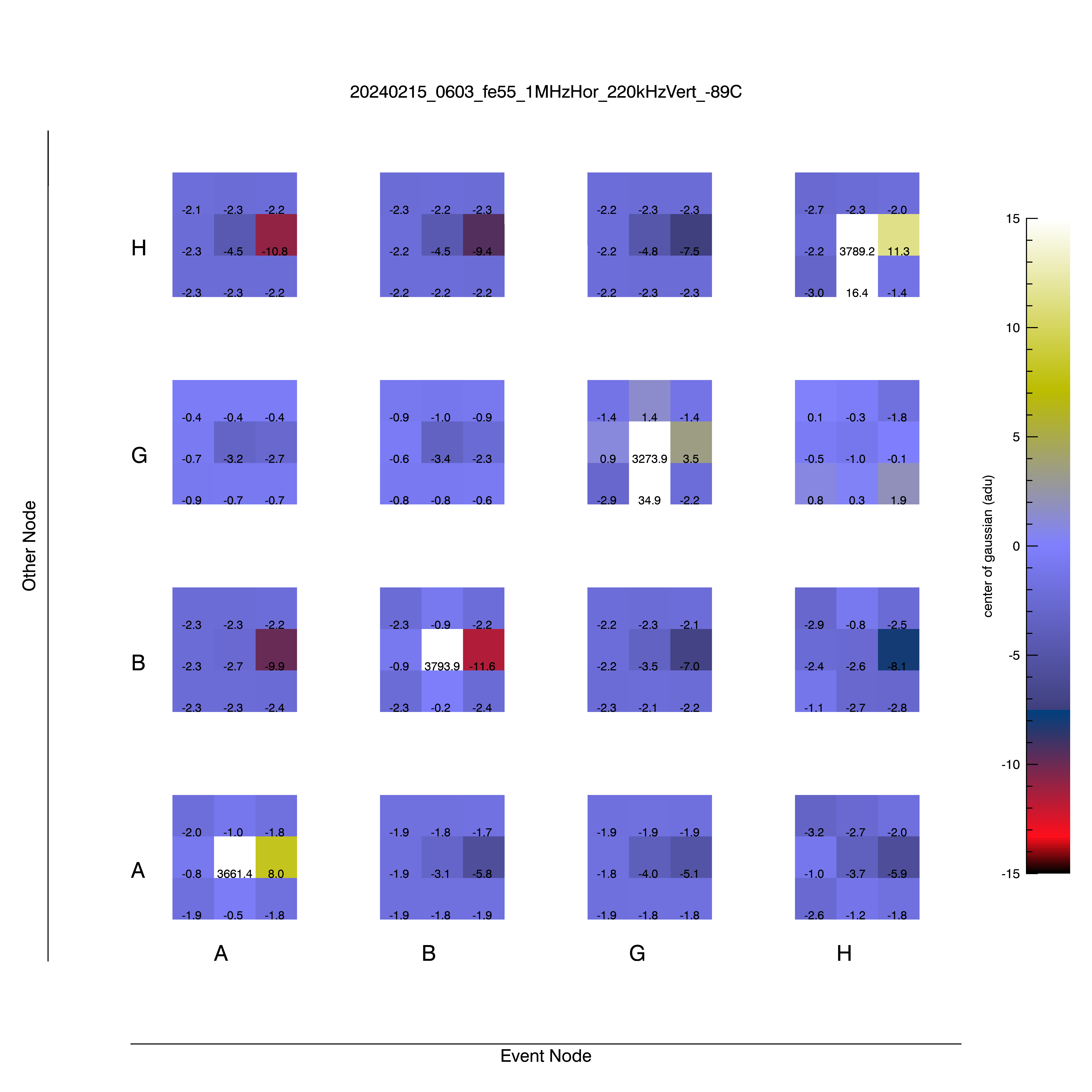}
    \caption{Signal amplitudes in the island surrounding $^{55}$Fe event and corresponding islands in the other sectors.}
    \label{fig:ccid89_crosstalk}
\end{figure}

Crosstalk between the CCD channels was measured by observing the signal levels in the other channels at the time when an X-ray event was registered in one of the sectors. The results are shown in the picture on Figure \ref{fig:ccid89_crosstalk} for the same 4 sectors of the CCD, named A, B, G, H.
Signal amplitudes are reported in islands of 3 x 3 pixels surrounding the central pixel of an X-ray event. The signal amplitudes corresponding to that same event in other sectors are shown in the pixel islands over the same column. Examining the signal values indicates that at the time of the event, amplitudes in the central pixels of other sectors drop by 1-2 ADUs relative to an undisturbed value. This constitutes approximately 0.06 \unit{\%} of the signal causing the disturbance, identifying such crosstalk to be a very small effect. Overall, the ASIC works very well in the multichannel regime and is ready to be used in real world applications.
\FloatBarrier
\subsection{Drain readout performance}
\label{sec:mcrc_dr_test}

The MCRC-V1 drain current input has a high speed p-MOSFET based active cascode (ACAS) followed by an adjustable current source (DRCS) and a current to voltage converter (I2V). The ACAS circuit is designed to isolate the very sensitive I2V negative input while also providing a stable voltage on the detector drain output. The adjustable current source provides the biasing current for the p-MOSFET drain and the ability to adjust the input current offset of the I2V amplifier. That is, when there is no signal present, the I2V input current is zeroed out to avoid saturation. The I2V is a two-stage operational amplifier with a transimpedance of 100 \unit{\kilo \ohm}. The drain readout input is an experimental circuit designed to be interfaced with detector output stages that use current modulation instead of voltage as the means of transmitting information. The advantage of such an architecture is its low impedance operation where the impact of parasitic capacitances is significantly diminished compared to the source follower, high impedance, based technology. The measured impedance at the ACAS input is $\sim$ 180 \unit{\ohm}. In theory, drain current based readouts can achieve higher speeds, lower noise, and better disturbance immunity. One such detector output is the so called Single electron Sensitive Readout (SiSeRO) \cite{tanmoySISERO1,tanmoySISERO2,tanmoySISERO3,tanmoySISERO4,Tanmoy2024} output. The SiSeRO output is composed of a p-type MOSFET transistor, which has two gates: an external and an internal gate. The external gate is used to set the operating point while the internal gate is used to modulate the drain-source current of the p-MOSFET output. The amplitude of the modulation is proportional to the charge that gets transferred to the internal gate from the last serial gate of the CCD serial register.

\begin{figure}[h]
\begin{center}
\begin{tabular}{c} 
\includegraphics[width=6.5in]{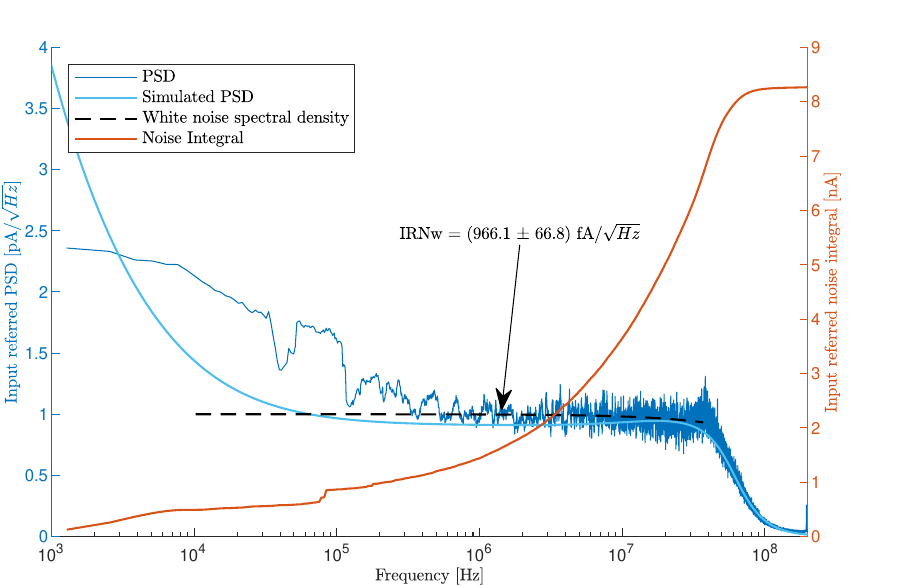}
\end{tabular}
\end{center} 
\caption{Drain current readout input referred noise spectral density.} 
\label{fig:mcrc_dr_noise}
\end{figure}

 Figure \ref{fig:mcrc_dr_noise} shows the simulated and measured noise spectral densities referred to the input of the current readout circuit. Both cases show good agreement in terms of the thermal noise floor with an input referred spectral noise density of $\sim$ 966 \unit{\femto A \per \sqrt{Hz}}.

\begin{figure}[h]
    \centering
    \begin{subfigure}{.48\textwidth}
    \includegraphics[width=\linewidth]{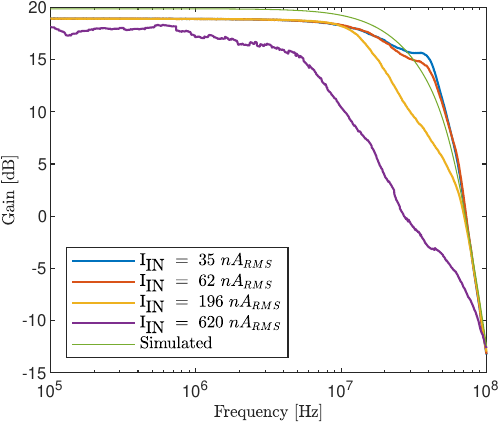}
    \caption{}
    \label{fig:mcrc_DR_gbw}
    \end{subfigure}
  \begin{subfigure}{.49\textwidth}
     \includegraphics[width=\linewidth]{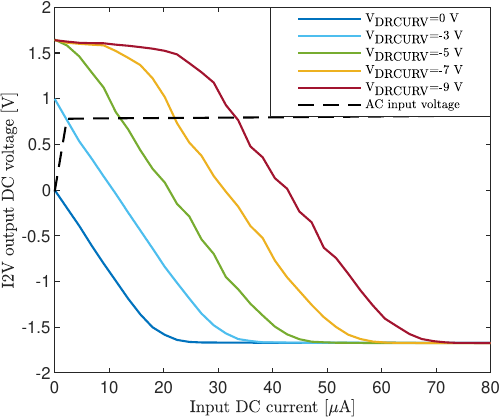}
    \caption{}
    \label{fig:mcrc_DR_curr}
  \end{subfigure}%
    \caption{(a) Gain scan of the current readout input circuit for different signal amplitudes. (b) I2V output voltage as a function of the input DC current at various current source voltages.}
    \label{fig:mcrc_DR_gbwandcurr}
\end{figure}

Figure \ref{fig:mcrc_DR_gbw} shows a gain scan of the current readout circuit at the output of the I2V. The small signal response shows good agreement with the simulated values, where the small signal bandwidth is $\sim$ 23 MHz. At large signals the response quickly degrades with the bandwidth dropping below 10 MHz. Figure \ref{fig:mcrc_DR_curr} shows the I2V output voltage as a function of input current for different source voltages of the DRCS. Adjusting the source voltage stirs the I2V response over the DR input current range. The input dynamic range of the circuit is limited by the input current range of the I2V, which is measured to be around 30 \unit{\micro A}.   

\FloatBarrier
\section{Conclusion and future upgrades}
\label{section:conclusions}
The MCRC-V1 prototype has been proven to provide a reliable substitute for discrete electronics implementations at a fraction of the power consumption and a significant reduction in real state area. The main findings from testing the MCRC-V1 prototype are summarized below:

\begin{itemize}
    \item The behavior of the MCRC-V1 is consistent with simulations. That is, the simulated optimal gain, bias, and reset settings closely match measurement results. 
    \item The high gain setting of 12 V/V provides a satisfactory dynamic range of $\sim$12 keV.
    \item Channel to channel variation in the MCRC-V1 is very low.
    \item There appears to be no significant effect on performance due to crosstalk (56 dB of channel-to-channel isolation).
    \item The noise, gain and spectral energy resolution performance of the MCRC-V1 is comparable to that of our best discrete electronics solutions but for only a fraction of the power and physical footprint.
    \item The MCRC-V1 noise contribution is $\sim$18 \unit{\%} of thermal noise to the total system noise. 
    \item Multichannel operation has been verified.
    \item Radiation tolerance has been demonstrated for a TID dose of 50 krad.
    \item Current readout input working with a noise floor of about 1 \unit{\pico A \per \sqrt{Hz}} and input impedance of roughly 180 \unit{\ohm}.  
    \item Optimal operating point settings:
    \begin{itemize}
        \item Source follower input current source voltage: -2.8 V.
        \item Preamplifier bias current: 75 \unit{\micro A}.
        \item Output buffer bias current: 45 \unit{\micro A}.
        \item MCRC-V1 reset mode: Line end, but pixel-by-pixel also possible.
        \item NCRC-V1 reset pulse width: 70 ns.
    \end{itemize}
\end{itemize}

Notwithstanding the overall success of the MCRC-V1 prototype, some bugs were discovered. These have been thoroughly analysed and understood. For the next iteration of the chip (MCRC-V2) we plan to apply all of the bug fixes known to date and also implement some additional features. One of these is an increase in channel count from 8 to 16 channels, to be optimally compatible with the upcoming CCID100 CCD detector, which will feature 16 p-jfet based outputs and will be the representative detector prototype for the AXIS flight hardware. Other new features include: the ability to enable/disable each output individually; an internal temperature sensor for better diagnostics; RG detector clock drivers to further reduce the local footprint area and power consumption, and alos provide better synchronization between the detector charge transfer window and readout ASIC reset times.

\acknowledgments 
 
This work has been supported by NASA APRA grants 80NSSC19K0499 and 80NSSC22K1921, SAT grant 80NSSC23K0211. We also thank the Kavli Institute of Particle Astrophysics and Cosmology for support via a KIPAC decadal funding grant.


\end{document}